\documentclass[twocolumn,showpacs,preprintnumbers,amsmath,amssymb]{revtex4}
\usepackage{graphicx}
\usepackage{bm}

\def\itmb{\begin{itemize}}
\def\itme{\end{itemize}}
\def\enmb{\begin{enumerate}}
\def\enme{\end{enumerate}}
\def\eqnb{\begin{equation}}
\def\eqne{\end{equation}}

\def\NPB{{Nucl. Phys.} {\bf B}}
\def\PLB{{Phys. Lett.} B}

\def\PRD{{Phys. Rev.} D}
\def\PRC{{Phys. Rev.} C}

\begin{document}
\newcommand{\ttbs}{\char'134}
\newcommand{\Slash}[1]{\ooalign{\hfil/\hfil\crcr$#1$}}

\title{Unqueched Kogut-Susskind quark propagator \\
in Lattice Landau Gauge QCD}
\author{Sadataka Furui}
\email{furui@umb.teikyo-u.ac.jp}
\homepage{http://albert.umb.teikyo-u.ac.jp/furui_lab/furuipbs.htm}
\affiliation{%
School of Science and Engineering, Teikyo University, 320-8551 Japan.
}%
\author{Hideo Nakajima}
\email{nakajima@is.utsunomiya-u.ac.jp}
 
\affiliation{
Department of Information Science, Utsunomiya University, 321-8585 Japan. 
}%

\date{\today}
\begin{abstract}
Quark propagators of the unquenched Kogut-Susskind(KS) fermion obtained from the  gauge configurations of the MILC collaboration are measured after Landau gauge fixing and using the Staple+Naik action. Presence of the $\bar q q$ condensates and $A^2$ condensates in the dynamical mass $M(q)$ and the quark wave function renormalization $Z_\psi(q^2)$ are investigated. We obtain the correlation of the renormalization factor of the running coupling taken at $\mu\sim 6$GeV and that of the quark wave function renormalization $Z_\psi(q^2)$ of the Staple+Naik action.
The mass function $\displaystyle M(q)$ is finite at $q=0$  and its chiral limit is $\sim 0.38$GeV. We compared the results corrected by the scale of the vertex renormalization and the tadpole renormalization with the corresponding values obtained by the Asqtad action without renormalization and observed good agreement.Implication of infrared finite  $Z_2(q)=1/Z_\psi(q^2)$  to the Kugo-Ojima confinement criterion is discussed.
\end{abstract}

\pacs{12.38.Gc, 12.38.Aw, 11.10.Gh, 11.15.Ha, 11.15.Tk, 11.30.Rd}
\maketitle
\section{Introduction}
The mechanism of dynamical chiral symmetry breaking and confinement is one of the most fundamental problem of hadron physics. The propagator of dynamical quarks in the infrared region provides information on dynamical chiral symmetry breaking and confinement.  In the previous paper\cite{FN04}, we measured gluon propagators and ghost propagators of unquenched gauge configurations obtained with quark actions of Wilson fermions (JLQCD/CP-PACS) and those of Kogut-Susskind(KS) fermions (MILC) in Landau gauge and observed that the configurations of the KS fermion are closer to the chiral limit than those of Wilson fermions.  

In the analysis of running coupling obtained from the gluon propagator and the ghost propagator, with use of the operator product expansion (OPE) of the Green function, we observed possible contribution of the quark condensates, $A^2$ condensates and gluon condensates in the configurations of the KS fermion\cite{FN04}. The quark propagator of quenched KS fermion was already measured in \cite{ABB}, and possible contribution of these condensates are reported. Unquenched KS fermion propagator of $20^3\times 64$ lattice (MILC$_c$) was measured in \cite{BHW}, but to distinguish the gluon condensates and the quark condensates, it is desirable to measure the quark propagator of larger lattice (MILC$_f$) and to compare with data of MILC$_c$.
We measure quark propagator of gauge configuration produced by using the Asqtad action 1) MILC$_c$ $20^3\times 64$, $\beta=6.76$ and 6.83 and 2) MILC$_f$ $28^3\times 96$, $\beta=7.09$ and 7.11.
@
The paper is organized as follows. In sect.2 we present kinematics of the staggered fermion on the lattice and in sect.3 we show the renormalization effects and in sect.4 numerical results of the mass function and the quark propagator are presented. Finally in sect.5 we give conclusion and discussion. In the appendix we show the algorithm of the inversion to get the quark propagator.

\section{Kinematics of the KS fermion}

The methods of writing the KS fermionic action on the lattice and deriving its propagator are presented in \cite{KMNP}. The KS fermion contains 16 flavor degrees of freedom and it is necessary to reduce the extra degrees of freedom by taking the 4th root of the fermion determinant. The MILC collaboration improved the fermionic action by fattening the gauge links, including Naik term and multiplying the tadpole renormalization factor so that the continuum limit can be approached on relatively small size of the lattice\cite{MILC1,MILC2}. Although there are no rigorous justification, there are indications that taking the 4th root does not yield serious problems.

In \cite{BHW}, a formulation of the KS fermion propagator calculation was presented and the numerical calculation was performed on MILC$_c$ ($20^3\times 64$ lattices) data\cite{bhlpwz}. In these papers, the momentum of quarks on the hypercubic lattice is defined by
\begin{equation}
p_\mu=\frac{2\pi n_\mu}{L}
\end{equation}
where $n_\mu=1,2,\cdots L/4$
and the 16 flavor degrees of freedom $\alpha_\mu=0,1$ where $\mu=1,\cdots,4$,
are expressed as
\begin{equation}
k_\mu=p_\mu+\pi\alpha_\mu
\end{equation}

In the Asqtad action, the link variable is modified by fattening 
\begin{eqnarray}
&&U^{fat}_\mu(x)= c_1 U_\mu(x)+\sum_\nu[ w_3 S^{(3)}_{\mu\nu}(x)\nonumber\\
&&+\sum_{\rho}(w_5 S^{(5)}_{\mu\nu\rho}(x)+\sum_\tau w_7 S^{(7)}_{\mu\nu\rho\tau}(x))]
\end{eqnarray}
where $S^{(3)}_{\mu\nu}$ is the staple contribution
\begin{equation}
S^{(3)}_{\mu\nu}(x)=U_\nu(x)U_\mu(x+\hat \nu)U^\dagger_\nu(x+\hat\mu) +h.c.
\end{equation}
$S^{(5)}$ and $S^{(7)}$ are 5-link and 7-link contribution, respectively.
In addition to fattening, so called Naik term and Lepage term are added.
The Naik term is a product of three link variables along one
direction\cite{MILC1}. 
\begin{equation}
U^{Naik}_\mu(x)=c_N U_\mu(x) U_\mu(x+\hat \mu) U_\mu(x+2\hat\mu)
\end{equation}
and the Lepage term is
\begin{eqnarray}
&&U^{Lepage}_\mu(x)=c_L U_\nu(x)U_\nu(x+\hat\nu)U_\mu(x+2\hat \nu)\nonumber\\
&&\times U^\dagger_\nu(x+\hat \nu+\hat \mu)U^\dagger_\nu(x+\hat\mu).
\end{eqnarray}
We compare the action including $S^{(3)}$ staple term and the Naik term only (Staple+Naik action), i.e. the same as that of \cite{OT}, Asq action and Asqtad action. 

The Staple+Naik action is
\begin{eqnarray}
&&S^{-1}_{\alpha\beta}(x,y)=\sum_{\mu=-4}^4 \eta_\mu(x)sign(\mu)\times[\nonumber\\
&&(c_1 U_\mu(x)+w_3\sum_{\nu\ne \mu}S_{\mu\nu}^{(3)}(x))\delta_{y,x+\hat\mu}\nonumber\\
&&+c_N U_\mu^{Naik}(x)\delta_{y,x+3\hat\mu} ]
\end{eqnarray}
where 
and $c_1=9/32$, $w_3=9/64$ and $c_N=-1/24$.

The Asq action contains $S^{(3)}, S^{(5)}, S^{(7)}$ and Lepage term and the Naik term  but no tadpole renormalization factor.
\begin{eqnarray}
&&S^{-1}_{\alpha\beta}(x,y)=\sum_{\mu=-4}^4 \eta_\mu(x)sign(\mu)\times[\nonumber\\
&&(c_1 U_\mu(x)+\sum_\nu w_3 S^{(3)}_{\mu\nu}(x)+\sum_{\rho}(w_5 S^{(5)}_{\mu\nu\rho}(x)\nonumber\\
&&+\sum_\tau w_7 S^{(7)}_{\mu\nu\rho\tau}(x))+c_L U^{Lepage}_\mu(x))\delta_{y,x+\mu}\nonumber\\
&&+c_N U^{Naik}_\mu(x)\delta_{y,x+3\mu}]
\end{eqnarray}
where $c_1=5/8$, $w_3=1/16$, $w_5=1/64$, $w_7=1/384$, $c_L=-1/16$ and $c_N=-1/24$.
 
The full Asqtad action contains the tadpole renormalization factor. Different from \cite{MILC2},  we do not absorb one power of $u_0$ into the quark mass in our calculation of the quark propagator, and use $c_1=5/8 u_0^{-1}$, $w_3=1/16 u_0^{-3}$, $w_5=1/64 u_0^{-5}$, $w_7=1/384 u_0^{-7}$, $c_L=-1/16 u_0^{-5}$ and $c_N=-1/24 u_0^{-3}$.

The inversion of the $S^{-1}_{\alpha\beta}(x,y)$ is performed via conjugate gradient method after preconditioning\cite{FFGLLK} as shown in the appendix.

We consider
\begin{eqnarray}
&&\Slash{D}_\mu(x,y)=\frac{1}{2}\sum_{\mu=1,\cdots, 4}\eta_\mu(x)
\nonumber\\
&\times&\{U^{fat}_\mu(x)\delta_{y,x+\mu}+U^{Naik}_\mu(x)\delta_{y,x+3\mu}\nonumber\\
&-&U^{fat}_\mu(x)^\dagger\delta_{y,x-\mu}-U^{Naik}_\mu(x)^\dagger\delta_{y,x-3\mu}\}
\end{eqnarray}
and solve for the bare valence quark mass $m_0$,
\begin{equation}
(\Slash{D}+m_0)\phi=\rho
\end{equation}
using the $3\times 3$ color matrix expression to the source term $\rho$. 

The propagator can be expressed in the form\cite{bhlpwz}
\begin{equation}
S(q)={\mathcal Z}_2(q)\frac{-i\Slash{q}+{\mathcal M}(q)}{q^2+{\mathcal M}(q)^2}
\end{equation}
Since $tr \gamma_\mu=0$, $tr$ over color and flavor yields
\begin{eqnarray}
tr S(q)&=&16N_c \frac{{\mathcal Z}_2(q){\mathcal M}(q)}{q^2+{\mathcal M}(q)^2}\nonumber\\
&=&16N_c{\mathcal B}(q)
\end{eqnarray}
On the other hand 
\begin{eqnarray}
tr (i\Slash{q} S)&=&tr {\mathcal Z}_2(q)\frac{q^2}{q^2+{\mathcal M}(q)^2}\nonumber\\
&=&16N_c q^2 \frac{{\mathcal Z}_2(q)}{q^2+{\mathcal M}(q)^2}\nonumber\\
&=&16N_c q^2{\mathcal A}(q)
\end{eqnarray}
The dynamical mass of the quark is 
\begin{equation}
{\mathcal M}(q)=\frac{{\mathcal B}(q)}{{\mathcal A}(q)}
\end{equation}
and the quark wave function renormalization is
\begin{equation}\label{z2}
{\mathcal Z}_2(q)=\frac{{\mathcal A}(q)^2q^2+{\mathcal B}(q)^2}{{\mathcal A}(q)}.
\end{equation}

We remark that our method of deriving ${\mathcal A}(q)$ and ${\mathcal B}(q)$ from the lattice data $S(q)$ is different from \cite{bhlpwz}, however the results are equivalent.

\section{Renormalization effects}

 In the continuum theory, the quark wavefunction renormalization $\displaystyle Z_\psi(q^2)$ is defined by the renormalized inverse quark propagator $S^{-1}(q)$ as\cite{pagels,Orsay1}
\begin{equation}
S^{-1}(q)=\delta_{ab}Z_\psi(q^2)(i\Slash{q}+M(q))
\end{equation}

On the lattice, renormalized operators must obey the same renormalization condition as the continuum theory. The quark field renormalization factor is usually defined by the amputated Green function of the vector current.
 
We define the colorless vector current vertex by using\cite{Orsay1}
\begin{equation}
G_\mu(q,p)=\int d^4 x d^4 ye^{iq\cdot y+ip\cdot x}\langle q(y)\bar q(x)\gamma_\mu q(x)\bar q(0)\rangle
\end{equation}
and
\begin{equation}
\Gamma_\mu(q,p)=S^{-1}(q)G_\mu(q,p)S^{-1}(p+q)
\end{equation}
The general expression of the vertex of the vector current is
\begin{eqnarray}
\Gamma_\mu(q)&=&\delta_{a,b}\{g_1(q^2)\gamma_\mu+ig_2(q^2)q_\mu+g_3(q^2)q_\mu\Slash{q}\nonumber\\
&&+ig_4(q^2)[\gamma_\mu,\Slash{q}]\}
\end{eqnarray}
where for 16 flavors
\begin{equation} g_1(q^2)=\frac{1}{48N_c}tr[\Gamma_\mu(q,p=0)(\gamma_\mu-q_\mu\frac{\Slash{q}}{q^2})].
\end{equation}

The Ward identity in the renormalized form tells
\begin{equation}
(\Gamma_R)_\mu(q)=-i\frac{\partial}{\partial q^\mu} S_R^{-1}(q)
\end{equation}
After multiplying both sides by $Z_\psi$, one obtains
\begin{equation}
Z_V\Gamma_\mu (q)=-i\frac{\partial}{\partial q^\mu} S^{-1}(q)
\end{equation}
where $Z_Vg_1(q^2)=Z_\psi(q^2)$ and $Z_V=1$ in the continuum limit
and  $Z_\psi(q^2)/g_1(q^2)=Z_V^{MOM}(a^{-1},q^2)$ is expected to be independent of $q$ in the limit of $a^{-1}\to \infty$.

In \cite{FN06}, we observed that the fluctuation of the ghost propagator in the quenched and unquenched simulations differ significantly. 

The renormalization of the Green function for the ghost-gluon coupling is
\begin{eqnarray}
\tilde G(q,g)&\to& \tilde z_3^{-1}G(q,g'),\nonumber\\
\Gamma_3(q,p,g)&\to& \tilde z_1\Gamma_3(q,p,g'),\nonumber\\
g&\to&g'=\tilde z_1 z_3^{-1/2}\tilde z_3^{-1}g
\end{eqnarray}

When a fermion of mass $m$ is coupled and the system is multiplicative renormalizable, the running coupling expressed by the dimensionless structure function of gluon $D$ and that of ghost $\tilde D$ and the vertex function $\Gamma_3$:
\begin{eqnarray}\label{ghost-gluon}
&&\bar g(\frac{p^2}{\Lambda^2},\frac{m^2}{\Lambda^2},g)=g\Gamma_3(\frac{q^2}{\Lambda^2},\frac{p^2}{\Lambda^2},\frac{m^2}{\Lambda^2},g)\nonumber\\
&&\times D(\frac{q^2}{\Lambda^2},\frac{m^2}{\Lambda^2},g)^{1/2}\times\tilde D(\frac{q^2}{\Lambda^2},\frac{m^2}{\Lambda^2},g)
\end{eqnarray}
is invariant under the change of the cut-off $\Lambda$ when $g$ is transformed to 
\begin{equation}\label{gluon-ghost}
g\to \tilde z_1^{-1}z_3^{1/2}\tilde z_3 g
\end{equation}

In the Dyson-Schwinger equation(DSE) aproach\cite{SHA,Blo1} the running coupling $g$ of the ghost-gluon coupling is 
\begin{equation}
g(q)=\tilde Z_1^{-1} Z_3^{1/2}(\mu^2,q^2) {\tilde Z}_3(\mu^2,q^2)g(\mu)\label{alphagluon}
\end{equation}
where $\tilde Z_1$ is the ghost vertex renormalization factor, which is taken to be 1 in pQCD. On the lattice, however, we took it as a scaling factor that adjust the r.h.s. of eq.(\ref{alphagluon}) at the $q=\mu\sim 6$GeV fits the value of pQCD\cite{FN04}.  The value of $1/{\tilde Z_1}^2$ is tabulated in Table\ref{z1fac}.
\begin{table}[htb]
\caption{The $1/\tilde Z_1^2$ factor of the unquenched SU(3).}\label{z1fac}
\begin{center}
\begin{tabular}{c|c|c|c|c}
 config.  & heavy  & light & $N_f$ &comments\\
\hline
MILC$_c$ &  1.49(11) & 1.43(10)& 2+1 &$\beta_{imp}=6.83, 6.76$\\
MILC$_f$ & 1.37(9) & 1.41(12)  & 2+1 &$\beta_{imp}=7.11, 7.09$\\
\hline
\end{tabular}
\end{center}
\end{table}

We observed that running coupling g in the rhs needed to be enhanced by $\tilde Z_1^{-1}$, so that it agrees with the pQCD. 

The running coupling of the quark gluon coupling is defined by the dimensionless structure function of gluon $D$ and that of quark $D_\psi$ and the vertex function $\Gamma_{3 \psi}$:
\begin{eqnarray}\label{quark-gluon}
&&\bar g(\frac{p^2}{\Lambda^2},\frac{m^2}{\Lambda^2},g)=g\Gamma_{3 \psi}(\frac{q^2}{\Lambda^2},\frac{p^2}{\Lambda^2},\frac{m^2}{\Lambda^2},g)\nonumber\\
&&\times D(\frac{q^2}{\Lambda^2},\frac{m^2}{\Lambda^2},g)^{1/2}\times D_\psi(\frac{q^2}{\Lambda^2},\frac{m^2}{\Lambda^2},g)
\end{eqnarray}
When the quark propagator is not calculated consistently using the simplified action as compared to the real action used in producing the gauge fields, one should define the normalization of the quark propagator $Z_\psi$ via other information.

Since at $\mu\sim 6$GeV, the quark wave function renormalization factor in perturbative QCD (pQCD) is close to 1, we define the scale of $Z_{\psi}(q^2)$ as 
\begin{equation}\label{Zpsiprime}
Z_{\psi}(q^2)=\frac{Z_Vg_1(\mu^2){\mathcal A}(q)}{{\mathcal A}(q)^2 q^2+{\mathcal B}(q)^2}
\end{equation}
where  $g_1(\mu^2)$ is defined from the running coupling in the following argument.
The function ${\mathcal B}(q)$ defines the quark inverse propagator $Z_\psi(q)$ and the dynamical mass $M(q)$ as
\begin{equation}
{\mathcal B}(q)={\mathcal A}(q){\mathcal M}(q)\propto \frac{1}{Z_\psi(q^2)}M(q).
\end{equation}

In the following, we define $M(q)={\mathcal M}(q)$ and renormalize the quark field as $\sqrt{Z_\psi(\mu^2)}\psi_{bare}=\psi_R$.
 
The Ward identity implies that the wavefunction renormalization can be defined via conserved vector current vertex.

The running coupling of quark-gluon coupling is calculated as
\begin{equation}
g(q)={Z_1^{\psi}}^{-1} Z_3^{1/2}(\mu^2,q^2) Z_2(\mu^2,q^2)g(\mu).\label{alphaquark}
\end{equation}

At the renormalization point $q=\mu$,  we fix $Z_2(\mu^2,\mu^2)=1$
and $Z_3(\mu^2,\mu^2)=1$, and thus
\begin{equation}
\tilde Z_1 {\tilde Z}_3^{-1}=Z_1^\psi Z_2^{-1}
\end{equation}
i.e. $\tilde Z_1=Z_1^\psi$. 

In \cite{Orsay1}, $Z_V$ is found to be almost constant for $q>5$GeV, and at $\mu\sim 6$GeV we identify  $Z_Vg_1(\mu^2)$ with $Z_1^\psi(\mu^2)$.
When quark wave function renormalization factor is measured using the Asqtad action, the running coupling given by eq.(\ref{alphaquark}) is expected to agree with the pQCD result in high momentum region. But in the calculation with Staple+Naik action, the wavefunction renormalization is to be modified by the tadpole renormalization factor $u_0$. Further, in the approximate calculation, the $Z_1^\psi$ would be modified as $\tilde Z_1$, and to give the consistent $g(q)$, additional correction to $Z_2(q)$ would be necessary. 

We define $Z_2(q)=1/Z_\psi(q^2)={\mathcal Z}_2(q)/Z_Vg_1(\mu^2)$ and identify $1/Z_Vg_1(\mu^2)$ for the Staple+Naik action as the average of the product of vertex renormalization factor and the tadpole renormalization factor i.e. $\tilde Z_1 u_0=1/1.38$ for MILC$_c$ and 1/1.36 for MILC$_f$. 

Thus, the renormalization of $Z_\psi$ on the lattice is defined by the renormalization of the running coupling on the lattice defined at $\mu\sim 6$GeV. 

\begin{figure}[htb]
\begin{center}
\includegraphics[width=7cm,angle=0,clip]{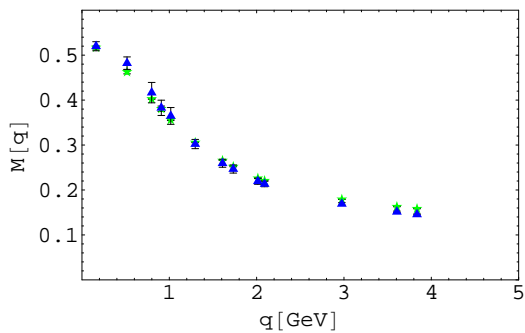}
\end{center}
\caption{The mass function $M(q)$ of Staple+Naik action(stars) and the Asqtad action(triangles) of MILC$_c$ with the bare quark mass $m_0=82.2$MeV.(Color online)}
\label{m_stp_atd}
\end{figure}

\begin{figure}[htb]
\begin{center}
\includegraphics[width=7cm,angle=0,clip]{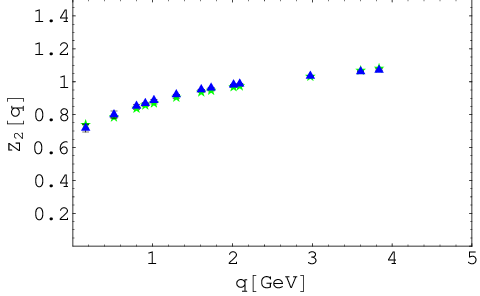}
\end{center}
\caption{The $Z_2(q)$ of Staple+Naik action(stars) and the Asqtad action(triangles) of MILC$_c$ with the bare quark mass $m_0=82.2$MeV.(Color online)}\label{z_stp_atd}
\end{figure}

In the case of Asq action, the renormalization factor ${\mathcal Z}_2(q)$ is suppressed by about 10\% and the mass function ${\mathcal M}(q)$ is enhanced by about 10\% as compared to the Staple+Naik.  By inclusion of the tadpole renormalization factor $u_0$ to the Asq action, the $Z_2(q)$ become consistent with those of Staple+Naik action with correction by $Z_Vg_1(\mu^2)$. In \cite{bhlpwz}, the scale of ${\mathcal Z}_2(q)$ calculated by the Asqtad action is fixed to 1 at $q=3$GeV, but we adopt the bare lattice data ${\mathcal Z}_2(q)=Z_2(q)$ and $M(q)$ of Asqtad action to compare with those of Staple+Naik with correction $Z_Vg_1(\mu^2)$.  Coincidence in the case of $m_0=82.2$MeV is shown in Figs.\ref{m_stp_atd} and \ref{z_stp_atd}. In the case of $m_0=11.5$MeV, the $M(q)$  and  $Z_2(q)$ of Staple+Naik in the infrared region are shifted from those of Asqtad slightly, but the difference is about 10\% at most, as shown in Figs.3 and 4, respectively.


\begin{figure}[htb]
\begin{center}
\includegraphics[width=7cm,angle=0,clip]{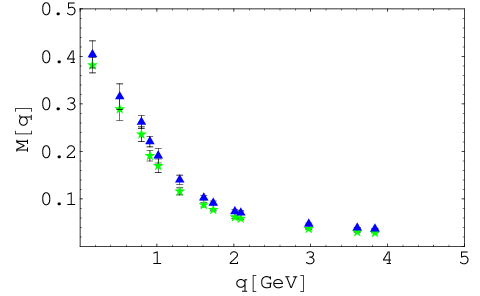}
\end{center}
\caption{The mass function $M(q)$ of Staple+Naik action(stars) and the Asqtad action(triangles) of MILC$_c$ with the bare quark mass $m_0=11.5$MeV.(Color online)}
\label{m_stp_atde}
\end{figure}
\begin{figure}[htb]
\begin{center}
\includegraphics[width=7cm,angle=0,clip]{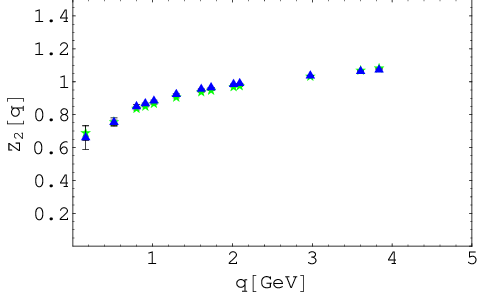}
\end{center}
\caption{The $Z_2(q)$ of Staple+Naik action(stars) and the Asqtad action(triangles) of MILC$_c$ with the bare quark mass $m_0=11.5$MeV.(Color online)}\label{z_stp_atde}
\end{figure}

\section{Analysis of $M(q)$ and $Z_\psi(q^2)$}
In perturbative QCD (pQCD), dynamical mass of a quark is expressed as\cite{ABB,Lane,Pol}
\begin{eqnarray}\label{massf}
M(q)&=&-\frac{4\pi^2 d_M\langle \bar q q\rangle_\mu [\log (q^2/\Lambda_{QCD}^2)]^{d_M-1}}{3q^2 [\log (\mu^2/\Lambda_{QCD}^2)]^{d_M}}\nonumber\\
&+&\frac{m(\mu^2)[\log (\mu^2/\Lambda_{QCD}^2)]^{d_M}}{[\log (q^2/\Lambda_{QCD}^2)]^{d_M}},
\end{eqnarray}
where $\displaystyle d_M=\frac{12}{33-2N_f}$. The second term is the contribution of the massive quark. 

In this analysis of the lattice data, the quark condensates $-\langle \bar q q(\mu)\rangle$ and $\Lambda_{QCD}$ are the fitting parameters. 
In the MILC$_f$ lattice, the bare masses are $0.0062/a=13.6$MeV and $0.0124/a=27.2$MeV for the $u-d$ quarks and $0.031/a=68.0$MeV for the $s-$quark.
In the MILC$_c$ lattice, the corresponfing masses are $0.007/a=11.5$MeV and $0.040/a=65.7$MeV for the $u-d$ quarks and $0.050/a=82.2$MeV for the $s-$quark.

The mass function eq.(\ref{massf}) is based on pQCD and cannot fit the data below 2GeV and we try the phenomenological fit\cite{SkW,BHW}
\begin{equation}
M(q)=\frac{c\Lambda^3}{q^2+\Lambda^2}+m_0
\end{equation} 
where $m_0$ is the bare quark mass. Parameters of $c$ and $\Lambda$ are summarized in TABLE \ref{massfit}. The fitting of the mass function of MILC$_c$ and that of MILC$_f$ are shown in Figs.5 and 6, respectively.

\begin{figure}[htb]
\begin{center}
\includegraphics[width=7cm,angle=0,clip]{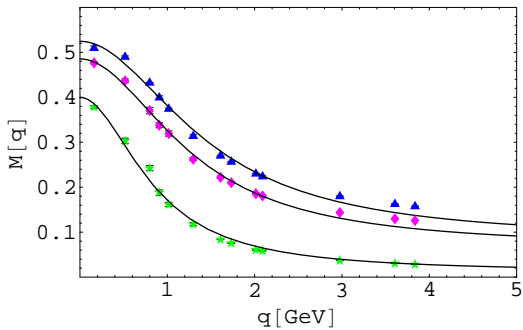}
\end{center}
\caption{The dynamical mass of the MILC$_c$ quark (Staple+Naik) with bear mass $m_0=11.5$MeV(stars), 65.7MeV(diamonds) and 82.2MeV(triangles) and the phenomenological fits. (Color online)}
\label{glpa}
\end{figure}
\begin{figure}[htb]
\begin{center}
\includegraphics[width=7cm,angle=0,clip]{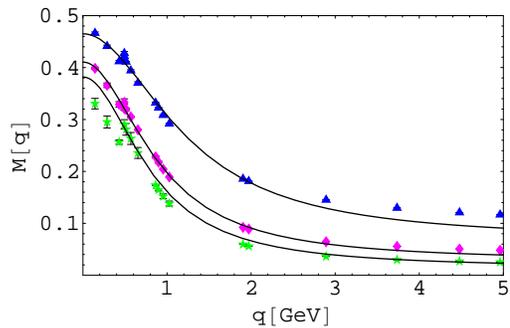}
\end{center}
\caption{Same as FIG. \ref{glpa} but MILC$_f$ quark mith bear mass $m_0=13.6$MeV(stars), 27.2MeV(diamonds) and 68.0MeV(triangles) and the phenomenological fits.(Color online) }%
\label{gld}
\end{figure}

\begin{table}[htb]
\caption{The parameters $c$ and $\Lambda$. }\label{massfit}
\begin{center}
\begin{tabular}{c|c|c|c|c}
$\beta_{imp}$ & $m_0$(MeV)     & $c$   &$\Lambda$(GeV) & $c\Lambda$(GeV)\\
\hline
6.76 & 11.5  & 0.44(1) & 0.87(2)  & 0.383 \\
     & 82.2  & 0.30(1)  & 1.45(2) & 0.431\\
6.83 & 65.7  & 0.33(1)  & 1.28(2) & 0.420 \\
     & 82.2  & 0.30(1)  & 1.45(2) & 0.431 \\
\hline
7.09 & 13.6  & 0.45(1) & 0.82(2) & 0.368\\
     & 68.0  &  0.30(1)  & 1.27(4) & 0.381 \\
7.11 & 27.2  &  0.43(1) & 0.89(2) & 0.383\\
     & 68.0  &  0.32(1)  & 1.23(2) & 0.397 \\
\hline
\end{tabular}
\end{center}
\end{table}

\begin{figure}[htb]
\begin{center}
\includegraphics[width=6cm,angle=0,clip]{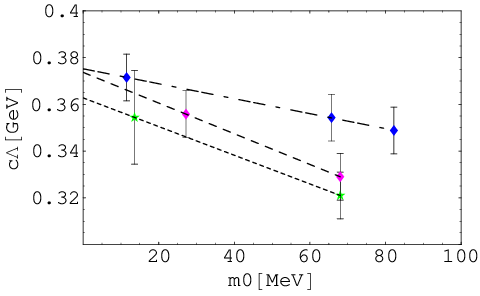}
\end{center}
\caption{The chiral symmetry breaking mass $c\Lambda=M(0)-m_0$ as a function of bare mass and its chiral limit. Dotted line is  the extrapolation of MILC$_f$ $\beta=7.09$, dashed line is $\beta=7.11$ and the dash-dotted line is that of MILC$_c$. (Color online)}\label{chiralmass}
\end{figure}

We observe that as the bare quark mass becomes heavy, $c$ becomes smaller but the product  $c\Lambda$ becomes larger.  Although the MILC$_c$ configurations of bare mass $m_0=82.2$MeV, with diffeernt $\beta$ agree within errors, the MILC$_f$ configurations of bare mass $m_0=68$MeV show dependence on $\beta$. The mass function of $\beta=7.09$ is smaller than that of $\beta=7.11$. In the case of $\beta=7.11$, the chiral limit $M(0)$ is consistent with that of MILC$_c$ and we find $M(0)$=0.37(1)GeV. However, in the case of $\beta=7.09$, $m_0=13.6$MeV,  the lowest three momentum points of $M(q)$ are systematically smaller than the other points.  The slope of the $\beta=7.09$ and that of 7.11 are almost the same. A preliminary analysis of $c\Lambda$ using the Asqtad action suggests that the data of $\beta=7.09$ and MILC$_c$ obtained by the Staple+Naik are underestimation, 
and the discrepancy between MILC$_c$ and MILC$_f$ of about 0.02GeV remain.
We expect that the $M(0)$ in the continuum limit would be about 0.38GeV, consistent with the value obtained by meromorphic parametrization of lattice data\cite{adfm}.  In an DSE approach\cite{bprt}, deviation of the mass function $M(0)$ from a linear function of $m(\zeta=19$GeV) where $\zeta$ defines the scale of the system is claimed. Although $m(\zeta=19$GeV) is not identical to $m_0$, $M(0)-m(\zeta=19$GeV) of the DSE increases near the chiral limit in contrast to our naive fitting shown in FIG.\ref{chiralmass}, which implies that the chiral symmetry effect monotonically decreases as the bare mass increases.

\begin{figure}[htb]
\begin{center}
\includegraphics[width=7cm,angle=0,clip]{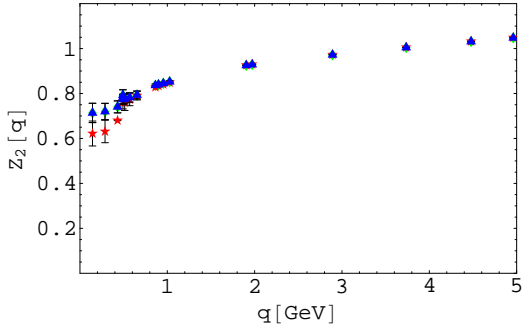}
\end{center}
\caption{The  $Z_2(q)$ of MILC$_f$ with bear mass $m_0=13.6$MeV(stars), 27.2MeV(diamonds) and 68MeV(triangles). (Color online)}
\label{z2f}
\end{figure}

\begin{figure}[htb]
\begin{center}
\includegraphics[width=7cm,angle=0,clip]{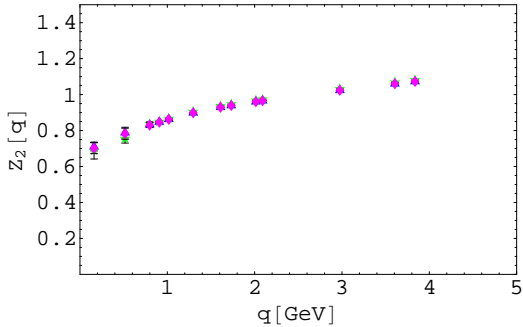}
\end{center}
\caption{Same as FIG. \ref{z2f} but quark of MILC$_c$ with bear mass $m_0=11.5$MeV(stars), 65.7MeV(diamonds) and 82.2MeV(triangles). The last two almost overlap. (Color online)}
\label{z2c}
\end{figure}

We show the lattice results of $Z_2(q)$ of MILC$_f$ and MILC$_c$ in Figs. \ref{z2f} and \ref{z2c}, respectively.
 
 The apparent difference in the formulae of \cite{bhlpwz} and our work are only in the expression and in fact they are equivalent.  The ${\mathcal Z}_2(q)$ agree with each other.

The  $A(q)$  defined as $\displaystyle S(q)=\frac{1}{A(q)\Slash{q}-B(q)}$ is parametrized in the operator product expansion approach as\cite{LO,ALSS}
\begin{eqnarray}
A(q)&=&1+\frac{\pi \alpha_s(\mu^2)\langle A^2\rangle_\mu}{N_c q^2}\nonumber\\
&-&\frac{\pi \alpha_s(\mu^2)\langle F^2\rangle_\mu}{3N_cq^4}+\frac{3\pi\alpha_s(\mu^2)\langle \bar q g \Slash{A} q\rangle_\mu}{4q^4}
\end{eqnarray}
where  $\langle F^2\rangle_\mu$ is the gluon condensate,  $\alpha_s(\mu^2)\langle A^2\rangle_\mu$ is the $A^2$ condensate and $\langle \bar q g \Slash{A} q\rangle_\mu$ is the mixed condensate, which are the fitting parameters.

Similar parametrization was done by \cite{ABB}, in which $\mu=2$GeV and $A(q)$ for $q>1$GeV is parametrized as
\begin{equation}\label{ABBfit}
A(q)=1+\frac{c_1}{q^2}+\frac{c_2}{q^4}
\end{equation}
In this parametrization, we obtained $c_1=0.25$GeV$^2$, $c_2=0.061$GeV$^4$ in the case of $m_0=27$MeV.   Comparing with data of quenched simulation\cite{ABB}($c_1=0.37\pm 0.06$GeV$^2$ and $c_2=-0.25\pm 0.04$GeV$^4$), $c_1$ is the same order but $c_2$ is smaller and have opposite sign. The result depends on the choice of the renormalization point $\mu$. 

The pQCD result of the wave function renormalization factor $Z_2(q)$ in $\overline{MS}$ scheme up to four loop is given by \cite{ChRe,chet}. Orsay group
expanded the $\alpha_s^{\overline{MS}}$ in terms of $\alpha_s^{MOM}(q)=\alpha$ and obtained $Z_\psi^{pert}(q^2)=1/Z_2(q)$ as a function of $\alpha$
and  fitted the $Z_\psi(q^2)$ of Wilson overlap fermion in the $\widetilde{MOM}$ scheme 
\begin{eqnarray}
&&(Z_\psi^{pert}(\mu^2))^{-1}S^{-1}(q)=(Z_\psi^{pert}(\mu^2))^{-1}S^{-1}_{pert}(q)\nonumber\\
&&+i\Slash{p} \frac{d(q^2/\mu^2,\alpha(\mu))}{q^2}\frac{\langle A^2\rangle_\mu}{4(N_c^2-1)}\delta_{ab}+\cdots\nonumber\\
&&=i\Slash{p}\delta_{ab}\left(\frac{Z_\psi^{pert}(q^2)}{Z_\psi^{pert}(\mu^2)}+\frac{d(q^2/\mu^2,\alpha(\mu))}{q^2}\frac{\langle A^2\rangle_\mu}{4(N_c^2-1)}\right)\nonumber\\
&&+\cdots
\end{eqnarray}
where $d(q^2/\mu^2,\alpha(\mu))$ is the solution of the renormalization group equation
\begin{equation}
\{(-\gamma_0+\gamma_{A^2}^{(0)})\frac{\alpha(\mu)}{4\pi}+\frac{d}{d\log \mu^2}\}d(q^2/\mu^2,\alpha(\mu))=0
\end{equation}
which can be written as
\begin{equation}
d(q^2/\mu^2,\alpha(\mu))=d(1,\alpha(q))\left(\frac{\alpha(\mu)}{\alpha(q)}\right)^{(-\gamma_0+\gamma_{A^2})/\beta_0}
\end{equation}

The anomalous dimension of $A^2$ in the lowest order is\cite{chet,DVS,VKAV} $\displaystyle \gamma_{A^2}^{(0)}=\frac{35}{4}-\frac{2}{3}n_f$ and $\gamma_0=0$.  We adopt $\langle A^2\rangle$ as a fitting parameter and calculate
\begin{eqnarray}
&&Z_\psi(q^2)=\frac{Z_Vg_1(\mu^2)}{{\mathcal Z}_2(q)}\nonumber\\
&&=Z_\psi^{pert}(q^2)
+\frac{\left(\frac{\alpha(\mu)}{\alpha(q)}\right)^{(-\gamma_0+\gamma_{A^2})/\beta_0}}{q^2} \frac{\langle A^2\rangle_\mu}{4(N_c^2-1)}{Z_\psi^{pert}(\mu^2)}\nonumber\\
&&+\frac{c_2}{q^4} \label{Orsayfit}
\end{eqnarray}
where $\alpha(q)$ are data calculated in the $\widetilde{MOM}$ scheme using the same MILC$_f$ gauge configuration\cite{FN04}.

We fitted the $Z_\psi(q^2)$ of MILC$_f$, $m=27.2$MeV data and choosing $\mu=2$GeV and $c_2=0$, obtained $\langle A^2\rangle_\mu=1.6(3)$GeV$^2$, which is compatible with the Orsay group data $2.4\pm 0.3$GeV$^2$ for the Wilson fermion.

\begin{figure}[htb]
\begin{center}
\includegraphics[width=7cm,angle=0,clip]{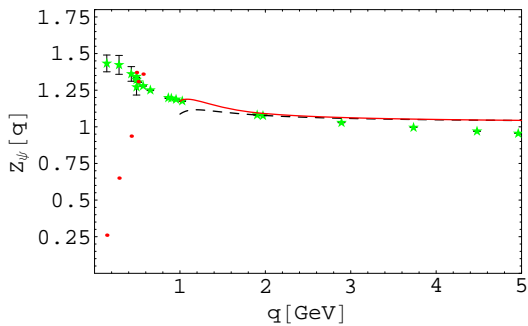}
\end{center}
\caption{The quark wave function renormalization $Z_\psi(q^2)$ of MILC$_f$ bear mass $m_0=27.2$MeV.
The solid line is the phenomenological fit as \cite{Orsay1} and the dashed line is the pQCD result. The points below 0.6GeV are $g_1(q^2)$ approximated by the running coupling obtained by the ghost-gluon vertex. (Color online)}\label{zze}
\end{figure}

In FIG.\ref{zze}, we show the lattice data of $\displaystyle Z_\psi(q^2)=\frac{Z_Vg_1(\mu^2)}{{\mathcal Z}_2(q)}$ from the Staple+Naik action with $Z_Vg_1(\mu^2)$=1.36 i.e. the average of the running coupling renormalization factor of MILC$_f$ in the previous analysis\cite{FN04}, and the phenomenological fit $Z_\psi^{fit}(q^2)$ by eq.(\ref{Orsayfit}).
The pQCD result $Z_\psi^{pert}(q^2)$ above $q=1$GeV is shown by the dashed line and the lattice data of $\alpha_s(q)$ below $q=0.5$GeV obtained by the ghost-gluon coupling are also plotted. The difference between $\alpha_s(q)$ and $Z_\psi(q^2)$ in the infrared implies that $Z_V$ is strongly momentum dependent in the infrared\cite{Orsay1}.

The suppression of $Z_\psi(q^2)$ in the infrared suggested by the behavior of $\alpha_s(q)$ could be an artefact. In the case of Wilson fermion\cite{blum}, the $Z_\psi(q^2)$ in the continuum limit is infrared finite.

\section{Conclusion and Discussion}
We measured the quark propagator of the MILC$_f$ lattice ($28^3\times 96$) of $\beta_{imp}=7.09$ and 7.11 and MILC$_c$ lattice ($20^3\times 64$) of $\beta_{imp}=6.76$ and 6.83 using the Staple+Naik action. After the renormalization using the information of the running coupling and the tadpole renormalization, we obtained a good agreement with the lattice bare data of the Asqtad action in the case of MILC$_c$.

We observed that the denominator of the quark renormalization factor 
\begin{equation}
{\mathcal Z}_2(q)=\frac{{\mathcal A}(q)^2q^2+{\mathcal B}(q)^2}{{\mathcal A}(q)}
\end{equation}
 is a steeply increasing function as $q$ approaches 0 which causes the infrared suppression. Although the function ${\mathcal B}(q)$ in the numerator is also an increasing function of $q$ as $q$ approaches 0, it is almost constant as the bare mass $m_0$ decreases, in contrast to ${\mathcal A}(q)$ which becomes larger as the bare mass $m_0$ decreases.  In the DS equation approach\cite{adfm},  $\sigma_v(q)$ and $\sigma_s(q)$ correspond to our ${\mathcal A}(q)$ and ${\mathcal B}(q)$, respectively. $\sigma_v(q)$ corresponds to $\displaystyle\frac{\Lambda_{QCD}}{q^2+m_0^2}$ and its $m_0$ dependence is qualitatively consistent with our lattice data. The dominant term of $\sigma_s(q)$ corresponds to $\displaystyle \frac{c\Lambda^3}{(q^2+m_0^2)(q^2+\Lambda^2)}$ and its $m_0$ dependence is weak in the $q\sim 0.1$GeV region consistent with our lattice data.

The infrared feature of $\displaystyle Z_\psi(q^2)=\frac{1}{Z_2(q)}$ is related to the Kugo-Ojima color confinement criterion:
\begin{equation}
1+u=\frac{Z_1}{Z_3}=\frac{\tilde Z_1}{\tilde Z_3}=\frac{Z_{1\psi}}{Z_2}=0
\end{equation}
i.e. the Kugo-Ojima parameter $u=-1$.

 In the quark sector, divergence of $Z_2(q)$ or vanishing of $Z_\psi(q^2)$ in the infrared is consistent with the Kugo-Ojima confinement criterion. Infrared finiteness of $Z_2$ implies that $Z_{1\psi}$ is infrared vanishing, and infrared finite $Z_\psi$ supports the idea that the running coupling freezes to a finite value in the infrared. This situation is similar to that of the gluon sector.  The infrared finite lattice data of $Z_3$ implies that $Z_1$ is also infrared vanishing. In this case running coupling defined by (\ref{alphagluon}) vanishes unless $Z_3$ is infrared vanishing. Definition of the running coupling from the triple gluon vertex is more ambiguous than the definition (\ref{alphagluon}) from the ghost-gluon vertex, and we expect that the fluctuation of the ghost propagator in the infrared induces the artefacts in the infrared running coupling.

 We compared the Staple+Naik action and the Asqtad action and found that the mass function and the quark propagator are consistent with each other.  They are consistent with the results of other group\cite{BHW,bhlpwz,pbhlwz}.

We confirmed $A^2$ condensates which was found by the Orsay group in the running coupling of quenched lattice simulation\cite{Orsay2} and observed in the unquenched lattice simulation\cite{FN04,Orsay1} also in the quark propagator. We observed also the  indication of the $\bar q q$ condensate.

The effect of dynamical chiral symmetry breaking is seen in the mass function $M(q)$. By extrapolation to the chiral limit we obtain $M(0)\sim 0.38$GeV, which is about 20\% larger than \cite{bhlpwz}, but consistent with DS equation approach\cite{adfm}.  An origin of the difference from \cite{bhlpwz} is their renormalization of the mass function $M(q)$ at 3GeV to a theoretical value, which we did not adopt. In the definition of the quark mass from the lattice data, one should take into account the mass renormalization by taking into account the lattice constant $a$ dependence of the propagator\cite{bglm,BL,blum,FL,McN}. 

Calculation of the quark propagator of MILC$_f$ using the Asqtad action is underway and the detailed comparison of the Asqtad action data and the DS equation approach will be published in the forthcoming paper.

\begin{acknowledgments}
We thank Tony Williams and Patrick Bowman for helpful informations.
Thanks are also due to MILC collaboration for the supply of their gauge configurations in the ILDG data base.
This work is supported by the KEK supercomputing project 05-128. H.N. is supported by the MEXT grant in aid of scientific research in priority area No.13135210.\end{acknowledgments}

\begin{center}
\appendix
\section{The Lattice calculation of the KS fermion propagator}
\end{center}
The Dirac gamma function of KS fermions are defined as\cite{BHW} 
\begin{equation}
(\bar\gamma_\mu)_{\alpha\beta}=(-1)^{\alpha_\mu}\delta_{\alpha+\zeta^{(\mu)},\beta}
\end{equation}
where 
\begin{equation}
\zeta^{(\mu)}_\nu=\left\{ \begin{array}{ll} 1 & \nu<\mu\\
                                            0 & {\rm otherwise}
\end{array}\right.
\end{equation}
In momentum space the inverse propagator is expressed as
\begin{equation}
S^{-1}_{\alpha\beta}(p,m)=i\sum_\mu(\bar \gamma_\mu)_{\alpha\beta}(\frac{9}{8}\sin p_\mu-\frac{1}{24}\sin 3p_\mu)+m\bar\delta_{\alpha\beta}
\end{equation}
where
\begin{equation}
\bar\delta_{\alpha\beta}=\prod_\mu \delta_{\alpha_\mu \beta_\mu} 
\end{equation}

The algorithm for calculating the propagator is as follows.
We define the operator devided by the bare mass of the dynamical quark $m$ as
\begin{equation}
{\bar M}=(I+\frac{1}{m}\Slash{D})
\end{equation}
and decompose even sites and odd sites,
\begin{equation}
{\bar M}=\left(\begin{array}{cc} I& \frac{1}{m}\Slash{D}_{oe}\\
                          \frac{1}{m}\Slash{D}_{eo} & I\end{array}\right)=I-L-U
\end{equation}
where 
\begin{equation}
L=\left(\begin{array}{cc} 0& 0\\
                          -\frac{1}{m}\Slash{D}_{eo} & 0\end{array}\right)
\end{equation}
and
\begin{equation}
U=\left(\begin{array}{cc} 0& -\frac{1}{m}\Slash{D}_{oe}\\
                          0 & 0\end{array}\right)
\end{equation}
Using the Eisenstat trick, we define
\begin{eqnarray}
\tilde {\bar M}&=&(I-L)^{-1}{\bar M}(I-U)^{-1}=(I+L)(I-U-L)(I+U)\nonumber\\
&=&I-LU=\left(\begin{array}{cc} I& 0\\
                          0 & I-(\frac{1}{m})^2\Slash{D}_{eo}\Slash{D}_{oe}\end{array}\right)
\end{eqnarray}

We note that $\tilde {\bar M}$ is hermitian and the conjugate gradient method and/or BiCGstab method are applicable. 

With use of the definition 
\begin{equation}
\frac{1}{m}\rho=\rho'=\left(\begin{array}{c}\rho_o'\\ \rho_e'\end{array}\right)
\end{equation}
and 
\begin{equation}
\phi=\left(\begin{array}{c}\phi_o'\\ \phi_e'\end{array}\right),
\end{equation}
we solve 
\begin{equation}
(I-\frac{1}{m^2}\Slash{D}_{eo}\Slash{D}_{oe})\tilde \phi_e=\rho_e'-\frac{1}{m}\Slash{D}\rho'_o
\end{equation}
with initial value $\tilde \phi_e^{(0)}=\tilde\rho_e'$. The odd site solution is $\tilde \phi_o=\rho_o'-\frac{\Slash{D}}{m}\tilde \phi_e$.

The decomposition into even sites and odd sites is done by multiplying
the projection operator $P_e$ and $P_o$, and the color source $\rho'$ is taken as a vector of $3\times 3$ matrices.

The convergence condition for the conjugate gradient iterations is
\begin{equation}
\frac{\| \tilde{\bar M}\phi-\rho \|}{\|\rho \|}<{\rm a\ few\ per\ cent\ at\ most}
\end{equation} 
where the used norm is maximum norm in the space of site, color and flavor, 
and the accuracy gets $10^{-1}$ higher if $L^2$ norm is used.

\end{document}